\documentclass[prl,twocolumn]{revtex4}

\usepackage{graphicx}
\usepackage{epsfig}
\usepackage{amsfonts}

\begin{document}
\title{Magnetoresistance of a 2D electron gas caused by electron interactions in the
 transition from the diffusive to the ballistic regime. }
\author{L. Li$^1$, Y.Y. Proskuryakov$^1$, A.K. Savchenko$^1$, E.H. Linfield$^2$,
D.A. Ritchie$^2$}
\address{$^1$ School of Physics, University of Exeter, Stocker Road,
Exeter, EX4 4QL, U.K.\\
$^2$ Cavendish Laboratory, University of Cambridge, Madingley
Road, Cambridge CB3 OHE, U.K.}

\begin{abstract}
On a high-mobility 2D electron gas we have observed, in strong
magnetic fields $(\omega_{c}\tau>1)$, a parabolic negative
magnetoresistance caused by electron-electron interactions in the
regime of $k_{B}T\tau/\hbar\sim1$, which is the transition from
the diffusive to the ballistic regime. From the temperature
dependence of this magnetoresistance the interaction correction to
the conductivity $\delta\sigma_{xx}^{ee}(T)$ is obtained in the
situation of a long-range fluctuation potential and strong
magnetic field. The results are compared with predictions of the
new theory of interaction-induced magnetoresistance.
\end{abstract}

\maketitle

Electron-electron interaction (EEI) corrections to the Drude
conductivity $\sigma_{0}$ of 2D systems have been intensively
studied over two decades.  These studies were based on the theory
of interactions in the diffusive regime, $k_{B}T\tau/\hbar<1$
\cite {Altshuler}. Physically this condition implies that the
effective interaction time, $\hbar/k_{B}T$, is larger than the
momentum relaxation time $\tau$ and therefore the two interacting
electrons experience scattering by many impurities. In the
ballistic regime, $k_{B}T\tau/\hbar>1$, electrons interact when
scattered by a single impurity.  A theory of the interaction
correction for such a case was only recently developed \cite
{Aleiner1}, and there have already been several experimental
attempts to apply it to the conductance of high-mobility (large
$\tau$) semiconductor structures
\cite{Yuri2,Shashkin,Pudalov,Vitkalov,Coleridge,Noh}. An essential
feature of this theory is that the impurities are treated as
point-like scatterers -- the condition which is not satisfied in
structures where the impurities are separated from the 2D channel
by an undoped spacer (unless the spacer is thick enough for the
background impurities to dominate the scattering). There is then a
question of how the interaction correction in the ballistic regime
manifests itself in a smooth fluctuation potential.

Introducing a long-range scattering potential is expected to
suppress the interaction correction in the ballistic regime
considered in \cite{Aleiner1}. This correction is caused by
electron back-scattering,  but in the case of a smooth potential
the backscattering is significantly reduced. However, as shown in
\cite{Rudin,Gornyi}, applying a strong magnetic field increases
the probability of an electron to return back and restores the
interaction correction.

Experimentally, the interaction correction
$\delta\sigma_{xx}^{ee}(T)$ is usually obtained from the
temperature dependence of the conductance, where it has to be
separated from the interference correction
$\delta\sigma_{xx}^{WL}(T)$ caused by the weak localisation (WL)
effect \cite{Lee}, as well as a possible classical contribution
from phonon scattering. It has been shown however that in the
diffusive regime there is an elegant method of detecting
$\delta\sigma_{xx}^{ee}(T)$ from the perpendicular
magnetoresistance \cite{Paalanen}. The interaction correction
 gives rise to the parabolic negative
magnetoresistance (NMR), expressed by the following relation at
$\omega_{c}\tau>1$:
\begin{equation}
\rho _{xx}=\frac 1{\sigma _0} + \frac 1{\sigma _0^2}(\mu
^2B^2)\delta \sigma _{xx}^{ee}(T),\label{1}
\end{equation}
where $\mu$ is the electron mobility. This relation is derived by
converting the conductivity tensor into the resistivity tensor and
using the fact that in the diffusive regime the Hall conductivity
is not affected by interactions: $\delta \sigma _{xy}^{ee}(T)=0$.
Another essential specific of the diffusive regime used in its
derivation is that $\delta\sigma_{xx}^{ee}(T)$ and
$\delta\sigma_{xy}^{ee}(T)$ are not changed with strong magnetic
field applied \cite{Houghton}.

There have been several experiments where this method of
extracting the interaction correction was used
\cite{Paalanen,Minkov,Choi}. However, apart from the experiment
\cite{Minkov} on low-mobility structures, these experiments were
in fact performed not in the diffusive but ballistic regime, where
Eq. 1 is not justified. The extracted quantity
$\delta\sigma_{xx}^{ee}(T)$ should be neither a logarithmic
correction in the diffusive regime \cite{Altshuler}, nor a linear
correction in the ballistic regime \cite{Aleiner1} derived for
classically small magnetic fields.

The new theory \cite {Gornyi} considers the interactions in the
ballistic and intermediate regimes in strong fields
$\omega_{c}\tau>1$.  It shows that interactions in systems with a
long-range potential will also produce a parabolic
magnetoresistance described by Eq. 1. According to this theory,
strong magnetic field not only restores the interaction
correction, but also provides the required condition for Eq. 1:
$\delta\sigma_{xy}^{ee}(T)/\sigma_{xy} < \delta \sigma
_{xx}^{ee}(T)/\sigma_{xx}$. The theory gives a distinct prediction
for the magnitude of the magnetoresistance $\delta \sigma
_{xx}^{ee}(T)$ in Eq. 1.  The aim of this work is to study the
magnetoresistance caused by electron-electron interactions in the
intermediate regime, in a structure with long-range fluctuation
potential, and to compare the results with the prediction of
\cite{Gornyi}.

The sample is a standard modulation doped n-GaAs heterostructure.
The doped layer is separated from the conducting channel by an
undoped spacer $d=20$ nm. The wafer has been mesa etched to a Hall
bar pattern. Measurements have been performed by a standard
4-terminal method with a current of 0.4 - 2 nA. The electron
mobility changes in the range
$0.42\times10^{5}-5.5\times10^{5}$cm$^{2}$/Vs with increasing
carrier density, which is lower than that of the samples in \cite
{Paalanen}. The range of electron densities is from
$0.46\times10^{11}$ cm$^{-2}$ to $2\times10^{11}$ cm$^{-2}$. The
parameter $k_{B}T\tau/\hbar$ in our experiment is varied from 0.04
to 3.8. (This value is 3.3-- 33 in \cite{Paalanen}, 1.7--5.6 in
\cite {Choi} and 0.004--0.18 in \cite {Minkov}).

The magnetoresistance of the 2DEG with $n=6.8\times
10^{10}$cm$^{-2}$ is shown in Fig. 1(a). To analyse the data in
terms of theory \cite {Gornyi}, we have to prove first that the
experimental conditions satisfy the theoretical approximations.
\textit{Firstly}, in the measured electron density range the
$k_{F}d$ value varies from 1.2 to 2.2, which proves that the
fluctuation potential, with the correlation length $d$, is indeed
long-range. This is further supported by an estimation of the
ratio of the momentum relaxation time to the quantum time found
from the magnitude of the Shubnikov--de Haas oscillations:
$\tau/\tau_{q}$ varies from 24 at the highest to 4 at the lowest
electron density. \textit{Secondly}, the magnetoresistance is
analysed in the magnetic field range $\omega_c\tau>1$.
\textit{Thirdly}, the gated structure is suitable for the study of
interaction effects as the gate is separated from the 2DEG by 70
nm,  while the separation between electrons is in the range from
13 nm to 26 nm and therefore interactions are not screened by the
metallic gate.

One can see in Fig. 1 that the negative magnetoresistance exhibits
a sharp change in small fields, followed by a parabolic dependence
in higher fields. The sharp change is caused by the WL effect
which is suppressed by magnetic field. We will analyse the
parabolic magnetoresistance in the range of fields well above the
`transport' magnetic field $B_{tr}=\hbar/4De\tau\sim0.013$T, to
make sure that the WL contribution to the negative
magnetoresistance is negligible. On the other hand, the magnetic
field should not be too large, in order to prevent the development
of the positive magnetoresistance caused by the Zeeman effect on
the interaction correction \cite{Altshuler}. This condition is
also satisfied in our experiments where  $g\mu_{B}B/k_{B}T\lesssim
1$. The contribution of this effect to the measured
mangetoresistance can be estimated, from the theory of
interactions in the diffusive regime \cite{Altshuler} and in the
ballistic regime with point scatterers \cite{Aleiner1}, as being
less than $1\%$ in the entire range of studied fields and thus can
be neglected.

After the initial rapid change, the magnetoresistance develops a
parabolic dependence, Fig. 1 (b). It is temperature dependent and
shows immediately the qualitative features of the model
\cite{Gornyi}, Fig. 1(c). A flat region can be seen at small
fields, which is a clear indication that the long-range potential
at small field suppresses the interaction correction which is then
restored by larger fields. In accordance with the theory, the flat
region is better seen at higher temperatures. Another feature is
seen in Fig. 1(d) - with increasing temperature the
magnetoresistance becomes positive (in the model this is the
result of the Hartree term becoming dominant over the exchange
term at higher temperatures).

Before proceeding with the analysis of the strength of the
magnetoresistance at different temperatures and extracting the
temperature dependent EEI correction, we would like to note that
there is also a classical contribution to the parabolic negative
magnetoresistance in high-mobility structures which can also be
seen in strong fields \cite{Dmitriev, Polyakov}.  This NMR
originates from the fact that the magnetic field makes it possible
for electrons to `cycle' around impurities and become localised,
and hence do not contribute to the conductance. The magnitude of
this effect is very different for short-range and long-range
fluctuation potentials. In the case of short range scatterers the
classical magnetoresistance has the form \cite
{Baskin,Bobylev,Dmitriev}:
\begin{equation}
\rho_{xx}=\rho_{0}\frac{1-p}{1+p^{2}/(\omega_{c}\tau)^{2}} ,
\label{2}
\end{equation}
where $p=exp(-2\pi/\omega_{c}\tau)$ is the fraction of cycling
electrons, and $\omega_{c}$ is the cyclotron frequency. Fig. 2
shows the comparison of the prediction by \cite{Dmitriev} with
experimental results at the intermediate electron densities
$n=5.7\times10^{10}$ cm$^{-2}$ and $n=9.0\times10^{10}$ cm$^{-2}$.
At high densities the magnitude of the classical effect expected
for short-range scatterers is much stronger than the measured
magnetoresistance. Furthermore, the direction of its change with
varying density (different $\tau$) would be opposite to the
experimental situation. This proves again that in the studied
samples we are dealing with long-range rather than short-range
scattering. It is the presence of a smooth potential which
significantly decreases the classical NMR \cite{Polyakov}, since
the `cycling' electron trajectories will now become the
trajectories `wandering' in the potential landscape. For the
moment we assume that the contribution of the classical
magnetoresistance is negligible and attribute all magnetorestance
to the interaction correction $\delta\sigma^{ee}_{xx}(T)$ (later
we will come back to this question).

We plot the resistivity as a function of $B^2$ and from the slope
of the straight line obtain $\delta \sigma _{xx}^{ee}(T)$. Fig. 3
shows the temperature dependence of $\delta \sigma _{xx}^{ee}$ at
different electron densities, where experimental points
concentrate around one curve. This curve becomes rather close to
the interaction correction in the exchange channel \cite{Gornyi}
if one makes a vertical shift of the theoretical curve by
$\Delta\sigma=-0.07 e^2/h$. We want to emphasise that apart from
this small shift there are no adjustable parameters involved in
the analysis. It is interesting to note that the interaction
correction $\delta\sigma_{xx}^{ee}(T)$ found from the temperature
dependence of the conductance at $B=0$ is defined with an accuracy
of a constant contributing to a renormalised value of the momentum
relaxation time.  In the method of quadratic magnetoresistance,
this constant does not contribute to the magnetoresistance and no
shift is allowed in comparing the results with the theory. We will
discuss below a possible physical origin of this additional,
temperature independent contribution to the quadratic
magnetoresistance.

The theoretical curve plotted in Fig. 3 is given by the following
expression that describes electron interactions in the exchange
channel:
\begin{equation}
\frac{\delta\rho_{xx}(B)}{\rho_{0}}=-\frac{(\omega_{c}\tau)^{2}}{\pi
k_{F}l}G_{F}(T\tau), \label{3}
\end{equation}
where $k_{F}$ is the Fermi wave number and $l$ is the mean free
path.  The function $G_{F}(x)$ has the asymptotes
$G_{F}(x\ll1)\simeq-lnx+const$ and
$G_{F}(x\gg1)\simeq(c_{0}/2)x^{-1/2}$, with $c_{0}\simeq0.276$.

Measurements at different electron densities have enabled us to
cover a broad range of the parameter $T\tau$ and detect the
specific features of this dependence. It shows a logarithmic
behaviour at small $T\tau$, followed by a rapid disappearance of
the interaction correction at higher temperatures.  The results
show good agreement with the expected `saturation' at $T\tau\sim
0.5$, which means that the turning effect of magnetic field is
reduced with increasing temperature and electrons have less chance
to return back in the long-range potential.

Note that if one naively compared the obtained dependence
$\sigma_{xx}^{ee}(T)$ with the one calculated in \cite{Aleiner1}
(for point-like scatterers in zero field), a striking difference
would be seen, Fig. 3 (dotted line). The latter has a linear
asymptote at $T\tau\gg 1$ and does not show any saturation.

It is important to emphasise that the comparison was made with the
contribution from the exchange channel only, however it is known
that there is another (Hartree) term in interactions controlled by
the Fermi liquid interaction parameter $F_{0}^{\sigma}$. Comparing
the total interaction correction (exchange plus Hartree
\cite{Gornyi}) with the experiment shows that the Hartree
contribution is much smaller than the exchange contribution.  It
can be seen in Fig. 3 that within the experimental error the
magnitude of the parameter $F_{0}^{\sigma}$ in our case cannot be
larger than 0.1--0.2. The value of $F_{0}^{\sigma}$ depends on the
parameter $r_s=1/(\pi n)^{1/2}a_B$, which is the ratio of the
Coulomb and kinetic energy of electrons. The value of the Fermi
liquid parameter is only known for $r_s<1$:
$F_{0}^{\sigma}=-\frac{1}{2
\pi}\frac{r_{s}}{\sqrt{2-r_{s}^{2}}}\ln(\frac{\sqrt{2}+\sqrt{2-r_{s}^{2}}}{\sqrt{2}-\sqrt{2-r_{s}^{2}}})$
\cite{Aleiner1}. We plot this value in Fig. 4, together with
results of two recent experiments where it was determined, in
different 2D structures, at large $r_s$. (The latter results were
obtained on systems with little effect of the smooth potential,
using analysis based on theory \cite{Aleiner1}.) The overall trend
on Fig. 4 indicates that in our range of $r_s=1.2-2.6$ it is
reasonable to expect the value of $F_{0}^{\sigma}$ to be
$\sim-0.15$.

Let us now return to the observed shift in Fig. 3. We believe that
it is caused by a contribution from the classical NMR
\cite{Polyakov}, known to be $T$-independent, which we completely
ignored earlier because the long-range potential suppresses the
mechanism described by Eq. 2. Quantitatively, the parabolic
classical negative magnetoresistance depends on the ratio of the
short- and long-range scattering times, $\tau_s/\tau_L$. For the
case of $\tau_{L}>\tau_{s}$, the magnetoresistance is given by the
following relation \cite{Polyakov}:
\begin{equation}
\delta\rho_{xx}/\rho_{0}=-\omega_{c}^{2}/\omega_{0}^{2} ,
\label{4}
\end{equation}
where $\omega_{0}=(2\pi n_{s})^{1/2}v_{F}(2l_{s}/l_{L})^{1/4}$,
with $n_{s}$ the concentration of the strong scatterers, $l_{s}$
and $l_{L}$ the mean free paths for strong and smooth potential
scatterers, respectively. Increasing the contribution of
long-range scattering (increasing this ratio) significantly
suppresses the classical magnetoresistance. We can estimate the
value of $\tau_L$ using the expression for $\tau$ in the case of
remote donor scattering \cite{Davies}:
$\frac{1}{\tau}=\frac{(\pi\hbar)n_{s}^{2D}}{8m(k_{F}d)^{3}}$,
where $n_{s}^{2D}$ is the concentration of scattering donors which
is approximately equal to the electron density. For short-range
scatterers the estimation of $\tau$ can be done using the relation
 $\frac{1}{\tau}=\frac{h a_B n_{s}^{3D}}{m(k_{F}a_B)^{3}}$ for
 background impurity scattering, assuming that $n_{s}^{3D}\sim 3\times10^{14}$
 cm$^{-3}$. For $n=9.0\times10^{10}$cm$^{-2}$ these estimations give
 $\tau_L\sim6$ ps and $\tau_s\sim10$ ps, which are close to the momentum
 relaxation time $\tau=6.8$ ps. An estimate using Eq. 4 shows that this effect can indeed
 account for the experimentally observed shift. A more accurate comparison is,
however, complicated
 due to uncertainties in the values of $\tau_s$ and $\tau_L$ and the fact that Eq. 4 is,
 strictly speaking, only valid for $\tau_L\gg\tau_s$.

In conclusion, we have studied the magnetoresistance of a
high-mobility 2D electron gas in a GaAs/AlGaAs heterostructure
where the electron scattering is determined by a long-range
fluctuation potential. In classically strong magnetic fields we
have observed negative magnetoresistance, which is parabolic and
temperature dependent.  We have shown that this magnetoresistance
originates from the electron-electron correction to the
conductance.  We have extracted this correction and demonstrated
that it is well described by the recent theory of interactions in
the regime which is intermediate between the diffusive and
ballistic regimes.

We are grateful to I. V. Gornyi and A. D. Mirlin for stimulating
discussions, and I. L. Aleiner for reading the manuscript and
valuable comments.  L.L. and Y.Y.P. are grateful to the ORS scheme
for the financial support.

 \newpage

Figure captions:
\\
Fig. 1. (a) Longitudinal resistivity versus magnetic field for
electron density $n=6.8\times10^{10}$cm$^{-2}$ at different
temperatures: $T=0.2$, 0.8, 1.2 K. (b) The same data presented as
a function of $B^{2}$. (c) Zoomed-in region of $\rho_{xx}(B)$ from
(a) at $T=1.2$ K, showing a flat region at small fields. (d)
$\rho_{xx}$ versus $B^{2}$ for another density,
$n=9\times10^{10}$cm$^{-2}$, at $T=0.4$, 1.2, 4.2 K, showing a
transition from negative to positive magnetoresistance.
\\
Fig. 2. Solid lines: Negative magnetoresistance for the momentum
relaxation time $\tau = 6.8$ ps and $\tau = 2.3$ ps. Dashed lines:
NMR introduced by the classical effect \cite{Dmitriev}, with $\tau
= 6.8$ and 2.3 ps.
\\
Fig. 3. Conductivity correction due to interactions, obtained
experimentally at different electron densities $n=0.46, 0.57,
0.68, 0.90, 1.2, 1.4, 2.0 \times10^{11}$ cm$^{-2}$ (different
symbols for different densities, circles correspond to the lowest
density). Solid line - theoretical prediction for the correction
due to exchange interaction, shifted by $-0.07e^{2}/h$. Dashed
line - theory for the total correction which includes exchange
(Fock) and Hartree interactions with Fermi-liquid parameter
$F_{0}^{\sigma}=-0.15$. Dotted line shows the result of the
interaction theory for point-like scatterers in the transition
from the diffusive to the ballistic regime (exchange term at zero
magnetic field). Inset: the same results presented in the
logarithmic scales.
\\
Fig. 4. The dependence of the Fermi liquid parameter on $r_{s}$.
Dashed box indicates an approximate range of $r_{s}$ and
$F_0^\sigma$ for the structures in this work. Open squares:
results from Ref.\cite{Yuri2} for hole gas in p-GaAs
heterostructures. Solid squares: results from Ref.\cite{Shashkin}
for electron gas in Si MOSFETs. Solid line is the theoretical
curve for small $r_s$ \cite{Aleiner1}.


\begin{references}
\bibitem{Altshuler} B. L. Altshuler and A. G. Aronov,
\textit{Electron-electron Interaction in Disordered Systems},
Edited by A. L. Efros and M. Pollak, North-Holland (1985).
\bibitem{Aleiner1} G. Zala, B. N. Narozhny, and I. L. Aleiner, Phys. Rev. B {\bf 64, }214204 (2001).
\bibitem{Yuri2} Y. Y. Proskuryakov, A. K. Savchenko, S. S. Safonov, M. Pepper, M. Y. Simmons, and
D. A. Richie, Cond-mat/0109261 (2001).
\bibitem{Shashkin} A. A. Shashkin, S. V. Kravchenko, V. T. Dolgopolov, T. M. Klapwijk,
Cond-mat/0111478 (2001).
\bibitem{Vitkalov} S. A. Vitkalov, K. James, B. N. Narozhny, M. P. Sarachik, and T. M. Klapwijk,
Cond-mat/0204566 (2002).
\bibitem{Pudalov} V. M. Pudalov, M. E. Gershenson, H.
Kojima, N. Butch, E. M. Dizhur, G. Brunthaler, A. Prinz, and G.
Bauer, Phys. Rev. Lett. {\bf 88,}196404(2002); V. M. Pudalov, M.
E. Genshenson, H. Kojima, G. Brunthaler, A. Prinz, and G. Bauer,
Cond-mat/0205449.
\bibitem{Coleridge} P. T. Coleridge, A. S. Sachrajda, and P. Zawadzki, Phys. Rev. B {\bf 65,} 125328 (2002).
\bibitem{Noh} H. Noh, M. P. Lilly, D. C. Tsui, J. A. Simmons, L. N. Pfeiffer, K. W. West,
Cond-mat/0206519 (2002).
\bibitem{Rudin} M. Rudin, I. L. Aleiner, and L. I. Glazman,
Phys. Rev. Lett. {\bf 78}, 709 (1997).
\bibitem{Gornyi} I. V. Gornyi  and A. D. Mirlin, Cond-mat/0207557 (2002).
\bibitem{Lee} P A Lee, and T V Ramakrishnan, Review of Modern Physics, \textbf{57}, 287 (1985).
\bibitem{Paalanen}  M. A. Paalanen, D. C. Tsui, J. C. M. Hwang, Phys. Rev. Lett. {\bf
51,} 2226 (1983).
\bibitem{Houghton}  A. Houghton, J. R. Senna, and S. C. Ying, Phys. Rev. B {\bf25,} 2196 (1982).
\bibitem{Minkov}  G. M. Minkov, O. E. Rut, A. V. Germanenko, A. A. Sherstobitov, V. I. Shashkin,
O. I. Khrykin, and V. M. Daniltsev, Phys.Rev. B {\bf 64, } 235327
(2001).
\bibitem{Choi} K. K. Choi, D. C. Tsui, S. C. Palmateer, Phys. Rev. B {\bf 33,} 8216
(1986).
\bibitem{Dmitriev} A. Dmitriev, M. Dyakonov and R. Jullien, Phys. Rev. B {\bf 64,
} 233321 (2001).
\bibitem{Polyakov} A. D. Mirlin, D. G. Polyakov, F. Evers, and P. W\"olfle,  Phys. Rev. Lett. {\bf 87,
} 126805 (2001).
\bibitem{Baskin} E. M. Baskin , L. N. Magarill, and M. V. Entin, Sov. Phys. JETP,
{\bf 48}, 365 (1978); E. M. Baskin and M. V. Entin, Physica B {\bf
249-251}, 805 (1998).
\bibitem{Bobylev} A. V. Bobylev, F. A. Maao, A. Hansen, and E. H. Hauge, Phys. Rev. Lett. {\bf 75,
}197 (1995); A. V. Bobylev, F. A. Maao, A. Hansen, and E. H.
Hauge, J. Stat. Phys. {\bf 87, } 1205 (1997);
\bibitem{Yuri} Y. Y. Proskuryakov, A. K. Savchenko, S. S. Safonov, M. Pepper, M. Y. Simmons, and D. A. Richie,
 Phys. Rev. Lett. {\bf 86,} 4895 (2001).
\bibitem{Davies} J. H. Davies, \textit{The Physics of
Low-dimensional Semiconductors}, Cambridge University Press
(1998).

\end{references}
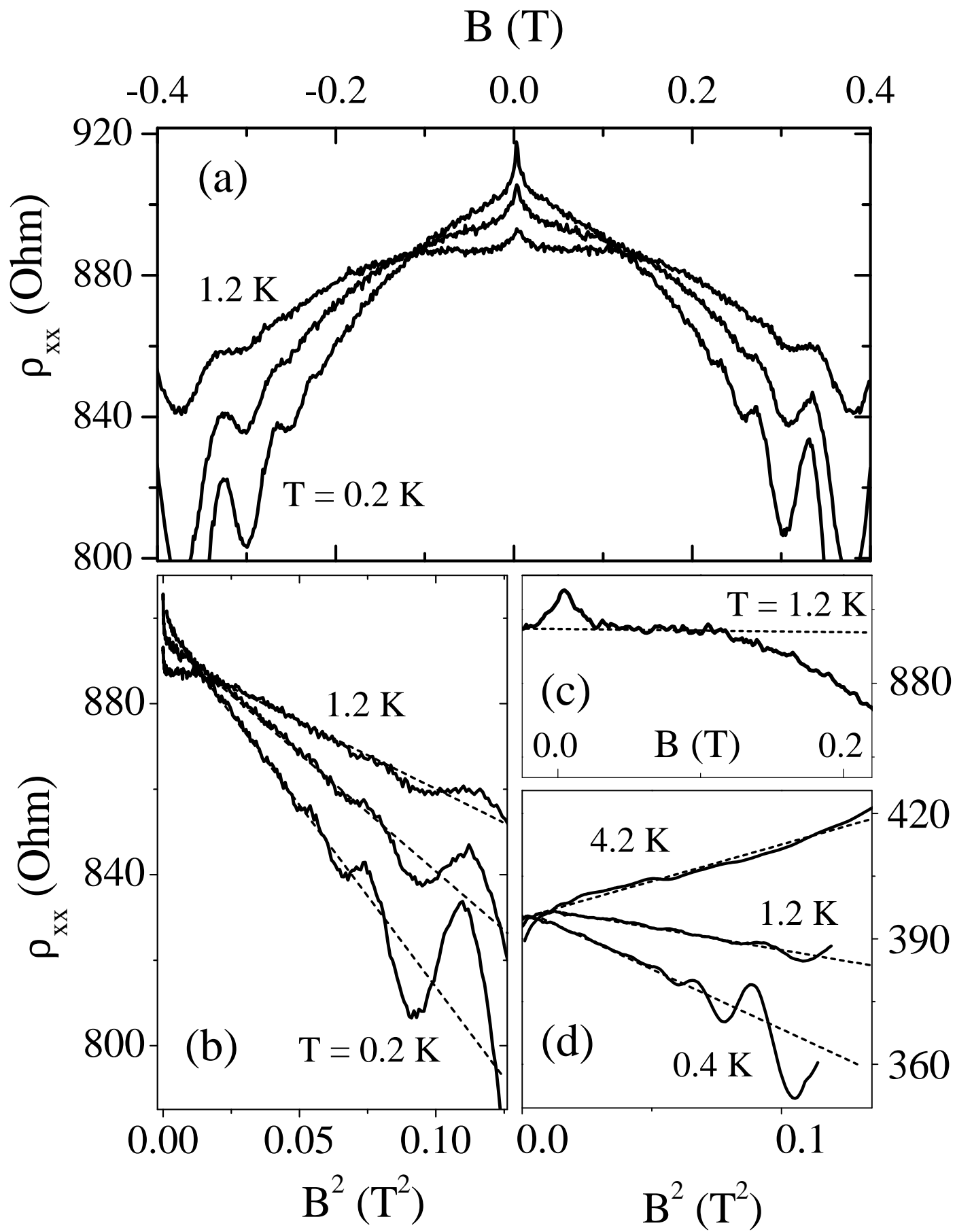
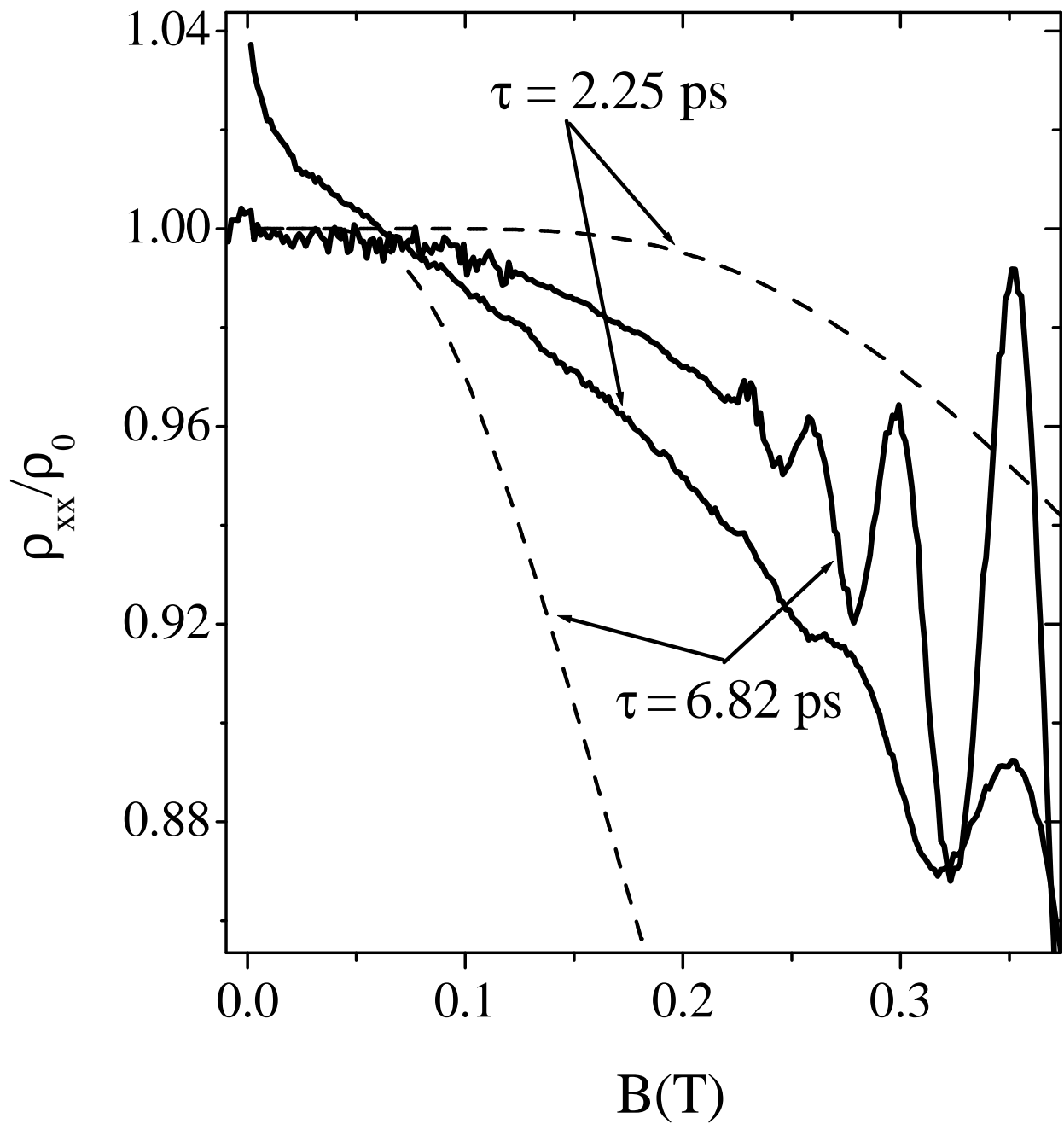
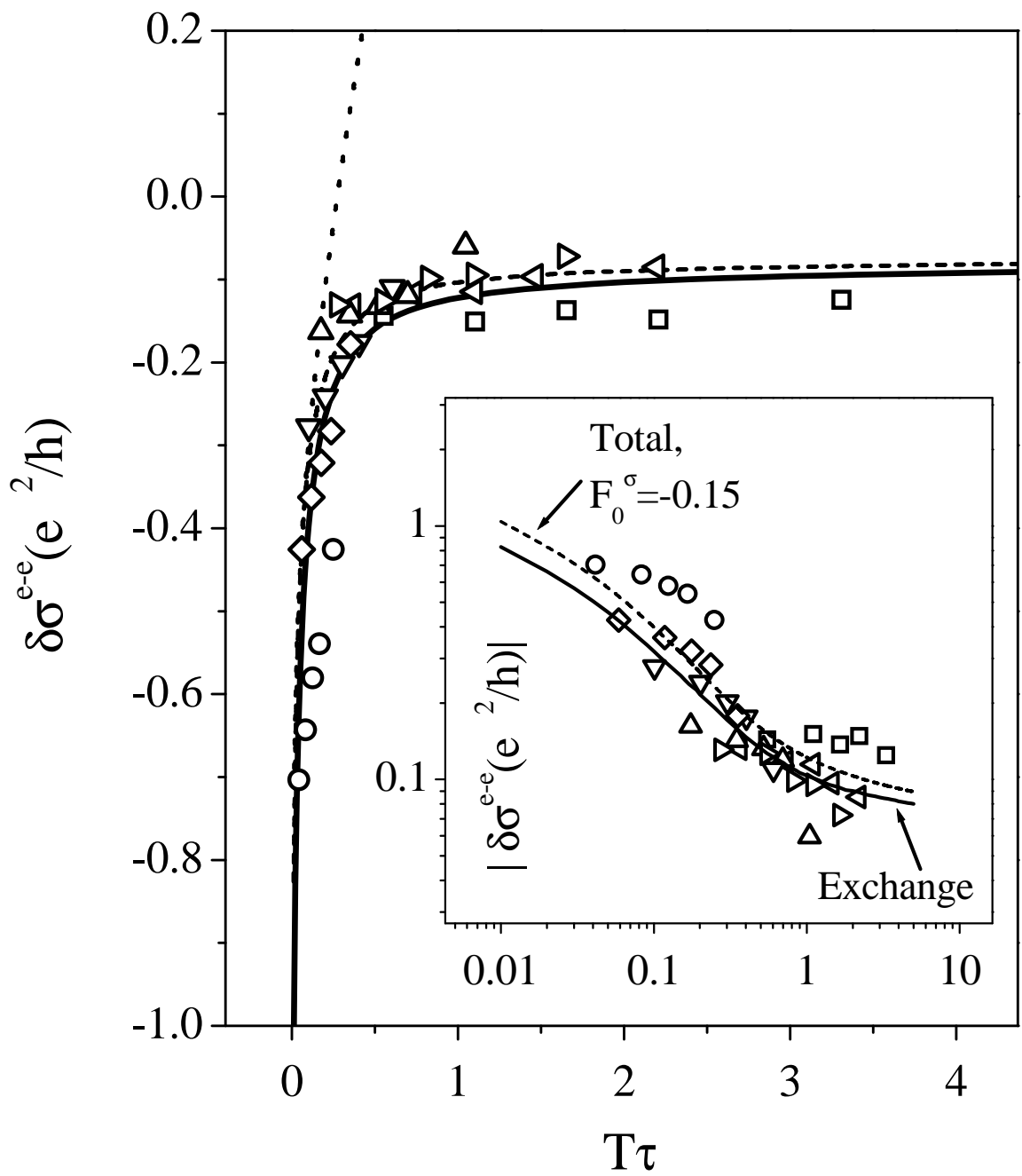
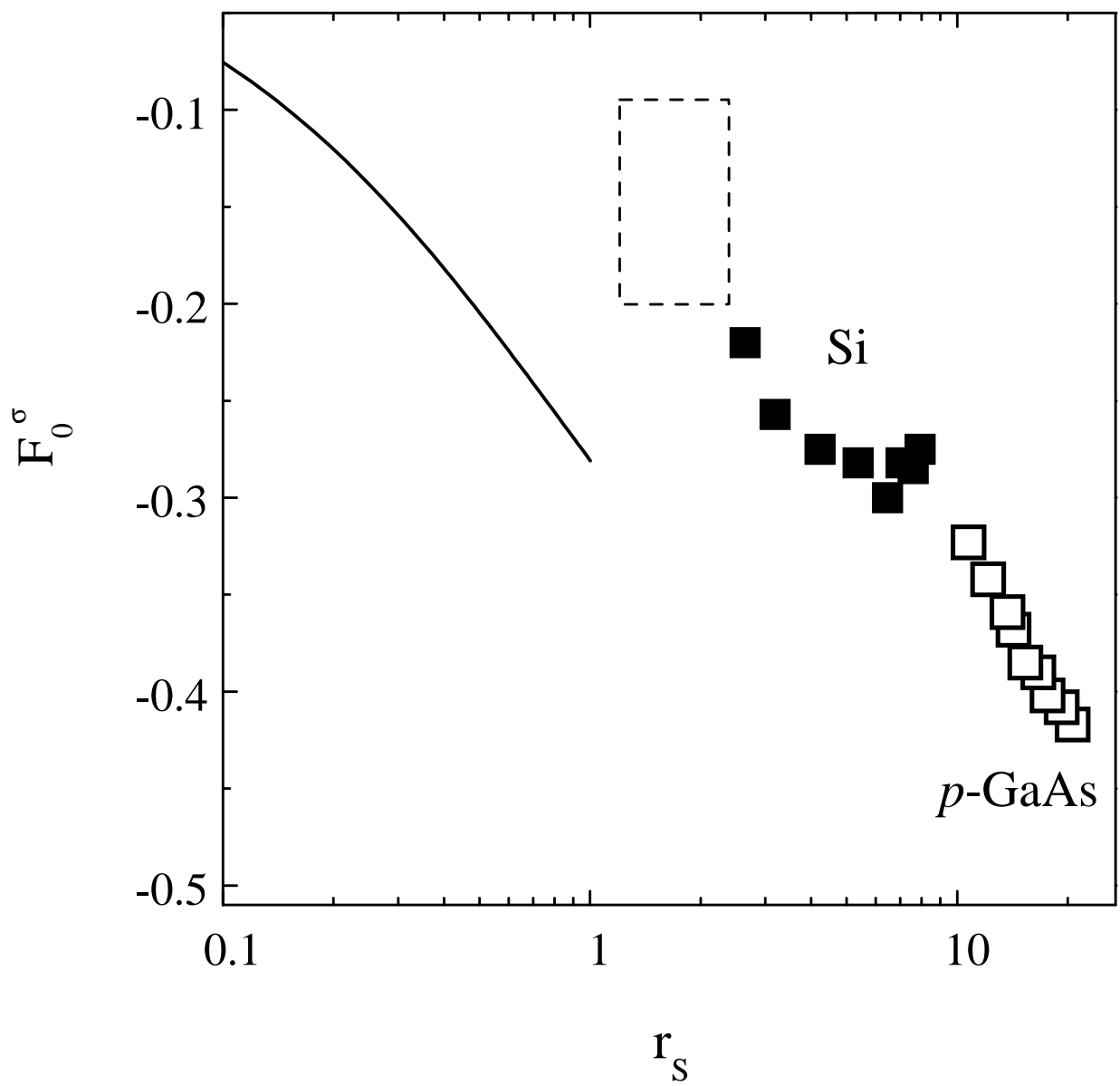
\end{document}